\documentclass{rspublic}
\begin{document}

\newcommand{\half}{\mbox{$\textstyle \frac{1}{2}$}}
\newcommand{\quat}{\mbox{$\textstyle \frac{1}{4}$}}

\title[Entropy and Temperature of a Quantum Carnot Engine]
{Entropy and Temperature of a \\ Quantum Carnot Engine}

\author[C.M. Bender, D.C. Brody and B.K. Meister]{
Carl M. Bender$^{*}$, Dorje C. Brody$^{\dagger}$ and Bernhard K.
Meister$^{\dagger}$ }

\affiliation{$*$Department of Physics,
Washington University, \\ St. Louis MO 63130, USA \\
$\dagger$Blackett Laboratory, Imperial College of Science,
Technology and Medicine, \\
London SW7 2BZ, UK \\}

\date{\today}
\maketitle
\input{psfig.sty}

\begin{abstract}
It is possible to extract work from a quantum-mechanical system
whose dynamics is governed by a time-dependent cyclic Hamiltonian.
An {\it energy bath} is required to operate such a quantum engine
in place of the heat bath used to run a conventional classical
thermodynamic heat engine. The effect of the energy bath is to
maintain the expectation value of the system Hamiltonian during an
isoenergetic process. It is shown that the existence of such a
bath leads to equilibrium quantum states that maximise the von
Neumann entropy. Quantum analogues of certain thermodynamic
relations are obtained that allow one to define the temperature of
the energy bath.  \vskip0.1cm
\begin{center}
{\footnotesize {\bf Keywords: quantum-mechanical engine; von
Neumann entropy; \\ thermodynamic equations of state} }
\end{center}
\end{abstract}

\section{Introduction}

A classical thermodynamic heat engine converts heat energy into
mechanical work by using a classical-mechanical system in which a
gas expands and pushes a piston in a cylinder. Such a heat engine
obtains its energy from a high-temperature heat reservoir. Some of
the energy taken from this reservoir is converted to mechanical
work. A heat engine is not perfectly efficient, so some of the
energy taken from the heat reservoir is not converted to
mechanical energy, but rather is transferred to a low-temperature
reservoir (Planck 1927).

A classical heat engine running between a high-temperature
reservoir and a low-temperature reservoir achieves maximum
efficiency if it is reversible. While it is impossible to
construct a working heat engine that is perfectly reversible, in
the early 19th century Carnot proposed a mathematical model of an
ideal heat engine that is not only reversible but also cyclic
(Carnot 1824). The Carnot engine consists of a cylinder of ideal
gas that is alternately placed in thermal contact with
high-temperature and low-temperature heat reservoirs whose
temperatures are $T_H$ and $T_C$, respectively.

Instead of a classical system of gas, we consider here a
quantum-mechanical system of consisting a single particle in
contact with a reservoir. The system is described by a
time-dependent cyclic Hamiltonian. The statistical ensemble of
many such identically-prepared systems that we consider here is
characterised by a density matrix. We assume that the system
interacts weakly with its environment. It is known that it is
possible to extract work from such a system (e.g., Geva $\&$
Kosloff 1994; Kosloff, {\it et al}. 2000).

In particular, if the evolution of the density matrix is
reversible, then we can construct a quantum-mechanical Carnot
engine (Bender, {\it et al}. 2000). The purpose of the present
paper is to investigate the properties of quantum Carnot engine.
This is of importance because it improves our understanding of the
relationship between thermodynamics and quantum mechanics, an area
of considerable interest in quantum theory (e.g., Leff $\&$ Rex
1990).

This paper is organised as follows: First, we explain the concept
of a quantum engine by generalising the two-state model considered
by Bender, {\it et al}. (2000) to an infinite-state square-well
model and derive equations of states for isoenergetic and
adiabatic processes. These results lead naturally to the quantum
analogue of the Clausius equality for a reversible cycle. We show
that, unlike the result in classical thermodynamics, the Clausius
relation obtained here is not based on the change of entropy.

We then introduce the von Neumann entropy and obtain the maximum
entropy state subject to isoenergetic requirements. We demonstrate
that the maximum von Neumann entropy is consistent with the
thermodynamic definition of entropy, as well as the equations of
states for quantum Carnot cycle. As a consequence, we are able to
determine the temperature of the energy bath directly from two of
the diagonal components of the density matrix. We conclude by
discussing several open problems.

\section{Quantum Carnot cycle}

We can construct a simple quantum engine using a single particle of
mass $m$ confined to an infinite one-dimensional square-well
potential whose volume (width) is $V$. For any fixed $V$ it is easy
to solve the time independent Schr\"odinger equation to determine
the energy spectrum of the system:
\begin{eqnarray}
E_n(V) = \frac{\pi^2\hbar^2n^2}{2mV^2}.
\label{eq:1}
\end{eqnarray}
We assume that the width of the well initially is $V=V_1$ and that
the initial energy of the system is a fixed constant $E_H$. The
initial state $\psi(x)$ of the system is a linear combination
$\psi(x)=\sum a_n\phi_n(x)$ of the energy eigenstates $\phi_n(x)$.
Thus,
\begin{eqnarray}
\sum_{n=1}^\infty p_n E_n(V_1) = E_H,
\label{eq:2}
\end{eqnarray}
where $p_{n}=|a_n|^2$, and $E_H$ is bounded below by
$E_H\geq E_1(V_1)$. Note that the pure state $\psi(x)$ characterises
a typical element of the ensemble, whose statistical property is
thus determined by the density matrix, given, in the energy
eigenstates, by $\rho_{mn}=p_{n}\delta_{mn}$.

Starting with the above initial configuration, we define a quantum
cycle as follows: First, the well expands {\it isoenergetically};
that is, the width of the well increases infinitely slowly while
the system is kept in contact with an energy bath. Note that the
quantum adiabatic theorem (Born $\&$ Fock 1928) states that if the
system were isolated during such an expansion, the system would
remain in its initial state. That is, the absolute values of the
expansion coefficients $|a_n|$ would remain constant. Thus, if the
system were isolated, the energy of the system (the expectation
value of the Hamiltonian) would decrease like $V^{-2}$.

However, during this expansion, we simultaneously pump energy into
the system in order to compensate this decrease of the energy.
Thus, the well, which can be viewed as a one-dimensional piston,
expands in such a way that the expectation value of the
Hamiltonian is held constant by exciting higher energy states.
This implies that, unlike the isothermal process in classical
thermodynamics, the temperature of the `bath' in an isoenergetic
process is in fact changing. The explicit volume dependence of
temperature is derived in Section 5.

During such an isoenergetic expansion, mechanical work is
done by the force (one-dimensional pressure) $P$ on the walls of the
well. The contribution to this force from the $n$th energy
eigenstate is
\begin{eqnarray}
f_n=\frac{\pi^2\hbar^2n^2}{mV^3}. \label{eq:3}
\end{eqnarray}
Hence the force $P$ is given by the expectation value $P=\sum p_n
f_n$. Using the relation $f_n=2E_n/V$ and Eq.~(\ref{eq:2}), we
obtain the equation of state during an isoenergetic process:
\begin{eqnarray}
PV=2E_H. \label{eq:4}
\end{eqnarray}
This result is identical to the
corresponding equation of state for an isothermal process of a
classical ideal gas, if we make the identification
$2E_{H}\leftrightarrow kT_H$. Note that the expansion coefficients
$a_n$ of the wave function change as a function of the width $V$
while the well expands isoenergetically from $V_1$ to $V_2$.

Second, the system expands {\it adiabatically}. During an adiabatic
process the eigenstates $\phi_n(x)$ change as a function of $V$, but
the values $|a_n|$ remain constant. Therefore, the expectation value
of the Hamiltonian $E=\sum p_nE_n(V)$ decreases during the process
because each $E_n$ decreases with increasing $V$ while all $p_n$ are
kept fixed. The force $P$ in this case is determined by
differentiating the energy $E$ with respect to $V$. Thus, the
equation of state during an adiabatic process for the square-well
potential is
\begin{eqnarray}
PV^3=2V_2^2E_H, \label{eq:5}
\end{eqnarray}
which is a quantum analogue of the corresponding equation for a
classical ideal gas. The system expands adiabatically until its
volume reaches $V=V_3$. At this point the expectation of the
Hamiltonian decreases to $E_C$. Because the squared coefficients
$p_n$ of the wave function remain constant during an adiabatic
process, the value of $E_C$ is given by $E_C=(V_2^2/V_3^2)E_H$.

Following the adiabatic expansion, the system is compressed
isoenergetically until $V=V_4$, with the expectation value of the
Hamiltonian fixed at $E_C$, and finally it is compressed
adiabatically until the width of the system returns to its initial
value $V=V_1$. This cycle is reversible, and we find that the
efficiency of the quantum engine is given by
\begin{eqnarray}
\eta=1-\frac{E_C}{E_H}, \label{eq:6}
\end{eqnarray}
a formula analogous to the classical thermodynamic result
$\eta=1-T_C/T_H$ of Carnot (1824).

\section{Quantum Clausius relations}

We have demonstrated above an example of a quantum system with a
slowly-changing time-dependent Hamiltonian from which work can be
extracted. The key
concept introduced here is an {\it energy bath} that maintains the
expectation value of the Hamiltonian. This idea can be developed
further to establish the quantum counterparts of the classical
thermodynamic relations. To begin with, let us first consider the
theorem of Clausius.

During an isoenergetic expansion, the amount of energy transferred to
the system to maintain the expectation of the Hamiltonian is
determined by the integral
\begin{eqnarray}
Q_H=\int_{V_1}^{V_2}dV\,P(V)=2E_H\ln(V_2/V_1), \label{eq:7}
\end{eqnarray}
where $P(V)=2E_{H}V^{-1}$. The amount of energy absorbed during an
isoenergetic compression can be determined similarly with the result
$Q_C=-2E_C \ln(V_3/V_4)$. Thus, for a closed, reversible cycle we
obtain the quantum Clausius equality
\begin{eqnarray}
\frac{Q_H}{E_H}+\frac{Q_C}{E_C}=0, \label{eq:8}
\end{eqnarray}
because $V_2/V_1=V_3/V_4$ for a closed cycle.

In classical thermodynamics the Clausius equality states that the
total change $\oint dS$ of entropy in a reversible cycle is zero,
where the differential of entropy $dS=dQ/T$ is the ratio of the
absorbed heat and the bath temperature. Therefore, the relation
(\ref{eq:8}) suggests that in quantum mechanics the entropy change
in an isoenergetic process might be given by $dQ/E$, the ratio of
the absorbed energy to the bath energy. However, as we show below,
this quantity does not determine the amount of entropy change, and
hence the quantum Clausius relation is not a condition for
entropy. Instead, it merely implies that the total energy
absorbed, for given bath energies, must vanish in a reversible
cycle. If the cycle is irreversible, then we have $\oint dQ/E<0$.

\section{Entropy maximisation}

To understand the change of entropy in a quantum Carnot cycle we
consider the von Neumann entropy
\begin{eqnarray}
S=-\sum_{n=1}^\infty p_n\ln p_n \label{eq:9}
\end{eqnarray}
associated with the density matrix $\rho_{mn}$,
expressed in terms of the energy eigenstates. Let us first consider
the isoenergetic process and introduce a scaling parameter $\lambda$
such that $V=\lambda V_1$. Then, during an isoenergetic process the
probability $p_n$ that the system is in the $n$th eigenstate changes
in $\lambda$, subject to the constraints
\begin{eqnarray}
\sum_{n=1}^\infty p_n = 1\quad{\rm and}\quad\sum_{n=1}^\infty
n^2p_n=\lambda^2, \label{eq:10}
\end{eqnarray}
where we have chosen units in which $\pi^2\hbar^2/(2m)=1$ and taken
the initial condition $E_H=E_1(V_1)$. There are infinitely many
combinations of $p_n$ satisfying these constraints, each of which
determines the entropy. Hence, quantum mechanically it appears that
any one of such states is allowed in an isoenergetic process.
However, there is a unique density matrix that satisfies the
thermodynamic requirement of the change of entropy, and this is the
state that maximises the von Neumann entropy.

To show this, we must determine the density matrix that maximises
the entropy subject to the constraints (\ref{eq:10}) and then
obtain the associated entropy $S$. We can perform this
maximisation directly without using Lagrange multipliers. The two
constraints in (\ref{eq:10}) imply that two of the diagonal
components, say $p_{k}$ and $p_{l}$, of the density matrix are
determined. Therefore, if we differentiate the constraints
(\ref{eq:10}) with respect to $p_{n}$, we find that
\begin{eqnarray}
\frac{\partial p_k}{\partial p_n}+ \frac{\partial p_l}{\partial
p_n}=-1\quad{\rm and}\quad k^2\frac{\partial p_k}{\partial
p_n}+l^2 \frac{\partial p_l}{\partial p_n}=-n^2. \label{eq:11}
\end{eqnarray}
Solving these linear equations, we obtain
\begin{eqnarray}
\frac{\partial p_k}{\partial p_n}=-\frac{n^2-l^2}{k^2-l^2}
\quad{\rm and}\quad
\frac{\partial p_l}{\partial p_n}=-\frac{n^2-k^2}{l^2-k^2}
\label{eq:12}
\end{eqnarray}
for all $k,l\neq n$. Substituting these expressions into
$\partial S/\partial p_n=0$ and solving for $p_n$, we get
\begin{eqnarray}
p_n = p_l\left(\frac{p_k}{p_l}\right)^{(n^2-l^2)/(k^2-l^2)},
\label{eq:13}
\end{eqnarray}
which is also valid for any choice of $k\neq l$.

Given the maximum entropy state (\ref{eq:13}), we can now reexpress
the constraints in the form
\begin{eqnarray}
\lambda^2 \sum_{n=1}^{\infty} \alpha^{n^2} =
\sum_{n=1}^{\infty} n^2 \alpha^{n^2} ,
\label{eq:14}
\end{eqnarray}
where we have defined $\alpha=(p_k/p_l)^{1/(k^2-l^2)}$. This is
obtained by substituting (\ref{eq:13}) in (\ref{eq:10}), solving
for $p_l$, and equating the resulting expressions. Similarly,
using the parameter $\alpha$, the state represented in
(\ref{eq:13}) can be expressed as
\begin{eqnarray}
p_n = \frac{\alpha^{n^2}}{\sum_{m=1}^{\infty}\alpha^{m^2}} .
\label{eq:15}
\end{eqnarray}
Therefore, the corresponding von Neumann entropy is
\begin{eqnarray}
S=(l^2-\lambda^2)\ln\alpha-\ln p_l.
\label{eq:16}
\end{eqnarray}
This is independent of $l$ because of the identity
\begin{eqnarray}
p_k\alpha^{-k^2}=p_l\alpha^{-l^2}. \label{eq:17}
\end{eqnarray}
Thus, given the width scale $\lambda=V/V_1$ of the potential well,
the relation (\ref{eq:14}) determines the value of $\alpha$, from
which we can determine both $p_n$ and $S$ as functions of the
single length scale parameter $\lambda$.

We can now analyse the entropy associated with the isoenergetic
process. We note first that the change of entropy associated with
(\ref{eq:16}) does not agree with the quantity $dQ/E$. More
specifically, we verify that only asymptotically for large values
of $\lambda$, that is, large energies, the two quantities are
proportional to each other. In particular, because the value of
$S$ in (\ref{eq:16}) is independent of the choices of $k$ and $l$,
we can set $k=2$ and $l=1$ without loss of generality. Then,
asymptotically we have $p_1\sim p_2$ for large $\lambda$.
Furthermore, the maximum entropy state $\{p_n\}$ satisfies $p_k>p_l$
for all $k<l$. Therefore, $\alpha\sim1$ and for large values of
$\lambda$ we use the asymptotic relation
\begin{eqnarray}
\sum_{n=1}^\infty(1-\epsilon)^{n^2}\sim
\frac{\sqrt{\pi}}{2\sqrt{\epsilon}}-
\frac{1}{2}\quad(\epsilon\to0^+) \label{eq:18}
\end{eqnarray}
to determine the entropy change. The result gives $\Delta
S\sim\ln\lambda$ when the width changes from $V_1$ to $V=\lambda
V_1$. On the other hand, from (\ref{eq:7}) we deduce that
$dQ/E=2\ln\lambda$ for any value of $\lambda\geq1$. This
establishes the asymptotic proportionality. However, $dQ/E$ does
not in general determine the entropy change.

\section{Equilibrium-state temperature}

Next, we show that the maximum von Neumann entropy (\ref{eq:16})
is consistent with thermodynamic considerations. Recall that an
isoenergetic process must proceed infinitely slowly in order that
the energy bath at any point during the process be in thermal
equilibrium. Thus, this process requires an infinite amount of
time to be realised. Then from the principles of statistical
mechanics (Schr\"odinger 1952) we deduce that, if we set $V_1=1$,
the temperature of the bath is given by
$T=-1/(\lambda^2\ln\alpha)$. This is because we can make the
identification $\alpha^{n^2}=\exp(-E_n/T)$ in Eq.~(\ref{eq:15}).
On the other hand, temperature is given in thermodynamics by the
formula
\begin{eqnarray}
\frac{1}{T} = \frac{dS}{dQ}, \label{eq:19}
\end{eqnarray}
where $dQ=PdV$. In the present consideration we have
$dQ=2\lambda^{-1}d\lambda$ from the equation of state $PV=2E_H$
and the relation $dQ=PdV$, while from (\ref{eq:16}) we find that
$dS/d\lambda=-2\lambda\ln\alpha$. Therefore, the thermodynamic
definition of temperature in (\ref{eq:19}) gives
\begin{eqnarray}
T=-\frac{1}{\lambda^2\ln\alpha}, \label{eq:20}
\end{eqnarray}
which agrees with the statistical mechanical consideration above.
We conclude therefore that the maximum von Neumann entropy
determines the density matrix of the system in an isoenergetic
process. Furthermore, the temperature of the energy bath can be
obtained from (\ref{eq:20}), where $\alpha$ is determined by any
two diagonal components $p_k$, $p_l$ of the density matrix. In
particular, the temperature of the bath changes continuously in
$\lambda$ during an isoenergetic process, even though the energy
of the system is held fixed.

We now consider the adiabatic process. In this case, there is no
net energy absorbed by the system, so the thermodynamic entropy
remains constant. On the other hand, because the diagonal
components $p_n$ of the density matrix are constant during an
adiabatic process, the von Neumann entropy (\ref{eq:9}) also
remains constant, in agreement with the thermodynamic
consideration.

These results can be expressed by the phase diagram in the
entropy-volume plane, as shown in figure~1. The diagram shows that
the thermodynamic constraint $V_2/V_1=V_3/V_4$ is satisfied by,
and only by, the maximum von Neumann entropy. Thus, we emphasise
that the quantum state corresponding to the maximum von Neumann
entropy, as obtained by microscopic analysis, is entirely
consistent with the thermodynamic equations of states.

\begin{figure}[t]
\label{} \psfig{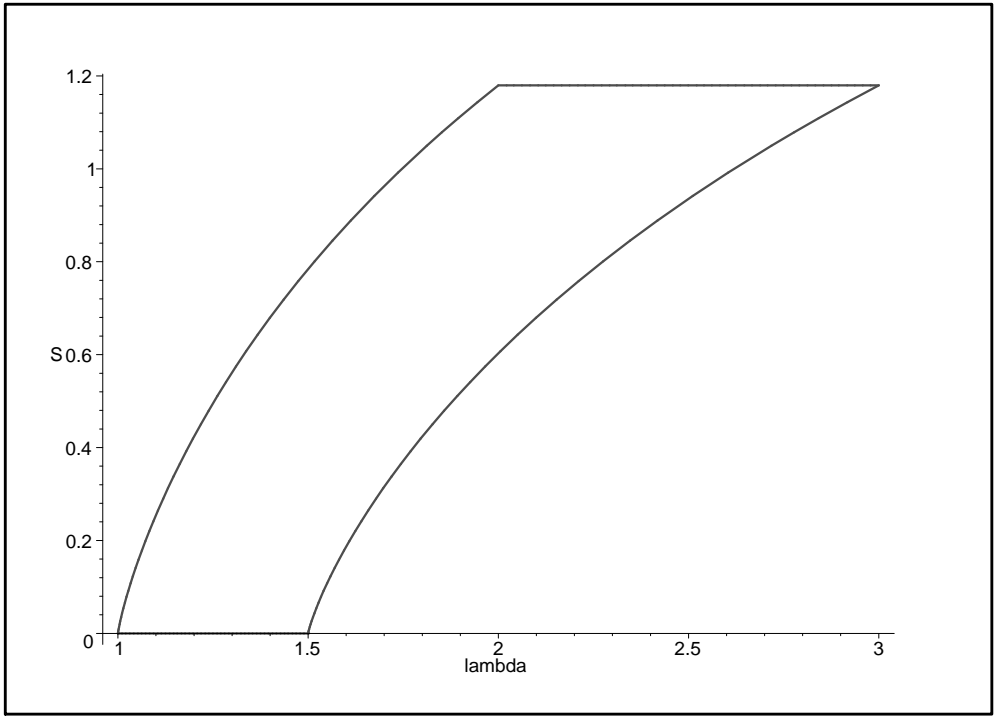}
\vspace{-2.5cm} \caption{{\it Phase diagram in the $S-\lambda$
plane}. The entropy of the system increases during an isoenergetic
expansion, satisfying the constraint $\langle E_n\rangle=E_H$,
then remains constant for an adiabatic expansion, decreases for
isoenergetic compression with the constraint $\langle
E_n\rangle=E_C$ to its initial value, and remains constant for
adiabatic compression. }
\end{figure}

\section{Discussion}

In the analysis above we have considered only the square-well
potential model, which can be interpreted as the quantum analogue of
the classical ideal gas. A natural extension of this work would be
to consider other Hamiltonians. For example, for a harmonic potential
whose energy eigenvalues are $E_n=(n+\frac{1}{2})\hbar\omega$, a
dimensional argument shows that the characteristic length scale is
given by $1/\sqrt{\omega}$. Therefore, the equations of states for
the harmonic potential become identical to those for the square-well
potential, although the volume dependence of the entropy is different
from (\ref{eq:16}). For other Hamiltonians, however, the equations of
states are not in general identical to those obtained here, and it
would be interesting to find explicit examples of other models
exhibiting nonideal behaviours.

Another important issue is to understand whether there is any role
played by the geometric phases, a concept that does not have an
analogue in the classical Carnot cycle. Although the quantum
Carnot engine is indeed cyclic in terms of $p_n$, the coefficients
$a_n$ of the wave function of any one of the representatives in
the ensemble inevitably pick up geometric phases as the system
goes through a cycle. Therefore, strictly speaking, the wave
function is not cyclic in the quantum Carnot cycle. The question
then is whether there is any physically observable evidence of the
geometric phase in the present context, and if so, how do we
represent that in the mixed state context (cf. Uhlmann 1996).

Finally, another interesting idea that should be studied is the
possibility of constructing a {\it cyclic} quantum engine (or a
quantum refrigerator by reversing the cycle) that requires only a
finite amount of time to complete. Note that each cycle of a
classical Carnot engine is infinitely long. However, in quantum
mechanics, it is known, for example, that by choosing a specific
form of time dependent Hamiltonians, it is possible to construct a
quantum-mechanical cycle in an essentially arbitrary short time
scale (Mielnik 1986). It would be of interest to determine if such
an idea can be applied to create a finite-time Carnot engine, or
whether such dynamics would be incompatible with the requirement
of maximising the von Neumann entropy. An analysis on the
performance of finite-time cycle in the presence of friction has
been studied in Feldmann $\&$ Kosloff (2000).

\vspace{0.5cm}

{\footnotesize We wish to express our gratitude to A.~Howie and
M.~F.~Parry for stimulating discussions. DCB gratefully
acknowledges financial support from The Royal Society. This work
was supported in part by the U.S.~Department of Energy.}

\end{document}